\documentclass[journal=jacsat,manuscript=article, doi=true]{achemso}
\setkeys{acs}{doi=true, maxauthors=0}
\usepackage{graphicx}
\usepackage{doi}
\usepackage[usenames,dvipsnames]{xcolor}
\usepackage[version=4]{mhchem}
\usepackage{amsmath}
\usepackage{siunitx}
\usepackage{caption}
\usepackage{amssymb}
\usepackage{pdfpages}
\usepackage{textgreek}
\usepackage[utf8]{inputenc}
\usepackage{etoolbox}
\makeatletter
\patchcmd{\@maketitle}{\begin{minipage}{\acs@maketitle@width}}{}{}{}
\patchcmd{\@maketitle}{\end{minipage}}{}{}{}
\makeatother

\title[Photodynamics of Isoxazole and Oxazole]{Heteroatom Position Controls Ultrafast Photodynamics of Oxazole and Isoxazole }

\author{Briony Downes-Ward}
\altaffiliation{These authors contributed equally to this work.}
\affiliation{Department of Chemistry, University of Missouri, Columbia, Missouri 65211, USA}
\email{brdfmg@missouri.edu}

\author{Paul Javed}
\affiliation{Department of Chemistry, Kansas State University, Manhattan, Kansas 66506, USA}
\altaffiliation{These authors contributed equally to this work.}

\author{Huynh V.~S.~Lam}
\affiliation{James R. Macdonald Laboratory, Department of Physics, Kansas State University, Manhattan, Kansas 66506, USA}

\author{Sajib K.~Saha}
\affiliation{James R. Macdonald Laboratory, Department of Physics, Kansas State University, Manhattan, Kansas 66506, USA}

\author{Surjendu Bhattacharyya}
\affiliation{Linac Coherent Light Source, SLAC National Accelerator Laboratory, Menlo Park, California 94025, USA}

\author{Martin Centurion}
\affiliation{Department of Physics and Astronomy, University of Nebraska-Lincoln, Lincoln, Nebraska 68588, USA}

\author{Xinxin Cheng}
\affiliation{Linac Coherent Light Source, SLAC National Accelerator Laboratory, Menlo Park, California 94025, USA}

\author{R.~Joel England}
\affiliation{Accelerator Directorate, SLAC National Accelerator Laboratory, Menlo Park, California 94025, USA}

\author{Casey Foley}
\affiliation{Department of Chemistry, University of Missouri, Columbia, Missouri 65211, USA}

\author{Smita Ganguly}
\affiliation{James R. Macdonald Laboratory, Department of Physics, Kansas State University, Manhattan, Kansas 66506, USA}

\author{Patrick L. Kramer}
\affiliation{Linac Coherent Light Source, SLAC National Accelerator Laboratory, Menlo Park, California 94025, USA}

\author{Jinxin Lang}
\affiliation{Department of Chemistry, University of Missouri, Columbia, Missouri 65211, USA}

\author{Randy Lemons}
\affiliation{Linac Coherent Light Source, SLAC National Accelerator Laboratory, Menlo Park, California 94025, USA}

\author{Ming-Fu Lin}
\affiliation{Linac Coherent Light Source, SLAC National Accelerator Laboratory, Menlo Park, California 94025, USA}

\author{Yusong Liu}
\affiliation{Linac Coherent Light Source, SLAC National Accelerator Laboratory, Menlo Park, California 94025, USA}

\author{Tu T.~Nguyen}
\affiliation{James R. Macdonald Laboratory, Department of Physics, Kansas State University, Manhattan, Kansas 66506, USA}

\author{Joao Pedro Figueira Nunes}
\affiliation{Diamond Light Source, Harwell Science and Innovation Campus, Didcot, OX11 0DE, UK}

\author{Chatura Perera}
\affiliation{Department of Chemistry, University of Missouri, Columbia, Missouri 65211, USA}

\author{Alexander H.~Reid}
\affiliation{Linac Coherent Light Source, SLAC National Accelerator Laboratory, Menlo Park, California 94025, USA}

\author{Ethan Ross}
\affiliation{Department of Chemistry, University of Missouri, Columbia, Missouri 65211, USA}

\author{Artem Rudenko}
\affiliation{James R. Macdonald Laboratory, Department of Physics, Kansas State University, Manhattan, Kansas 66506, USA}

\author{Xiaozhe Shen}
\affiliation{Accelerator Directorate, SLAC National Accelerator Laboratory, Menlo Park, California 94025, USA}

\author{John Searles}
\affiliation{James R. Macdonald Laboratory, Department of Physics, Kansas State University, Manhattan, Kansas 66506, USA}

\author{Anbu Selvam Venkatachalam}
\affiliation{James R. Macdonald Laboratory, Department of Physics, Kansas State University, Manhattan, Kansas 66506, USA}

\author{Enliang Wang}
\affiliation{Hefei National Research Center for Physical Sciences at the Microscale and Department of Modern Physics, University of Science and Technology of China, Hefei 230026, China}

\author{Stephen P.~Weathersby}
\affiliation{Accelerator Directorate, SLAC National Accelerator Laboratory, Menlo Park, California 94025, USA}

\author{Vinod Kumarappan}
\affiliation{James R. Macdonald Laboratory, Department of Physics, Kansas State University, Manhattan, Kansas 66506, USA}

\author{Arthur G. Suits}
\affiliation{Department of Chemistry, University of Missouri, Columbia, Missouri 65211, USA}

\author{Christine Aikens}
\affiliation{Department of Chemistry, Kansas State University, Manhattan, Kansas 66506, USA}
\email{cmaikens@ksu.edu}

\author{Daniel Rolles}
\affiliation{James R. Macdonald Laboratory, Department of Physics, Kansas State University, Manhattan, Kansas 66506, USA}
\email{rolles@ksu.edu}

\begin{document}

\begin{tocentry}
\includegraphics{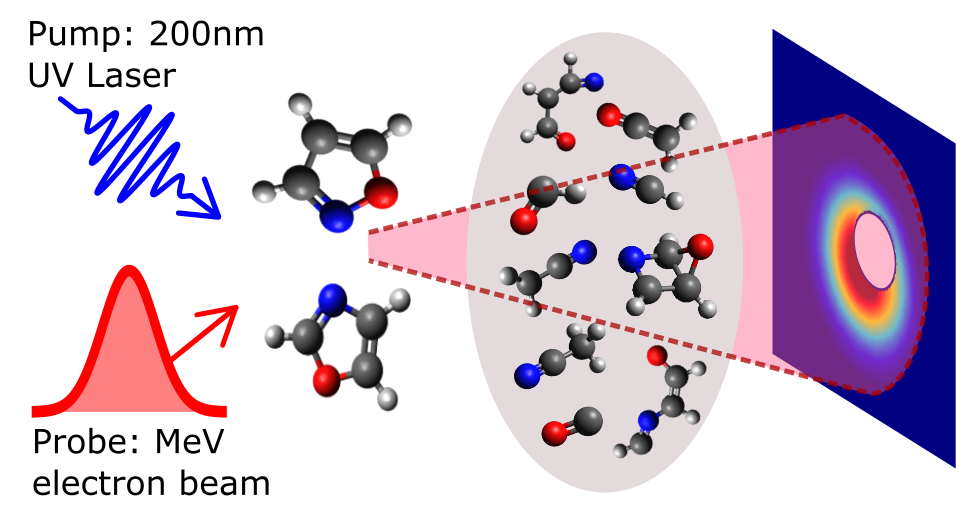}
\end{tocentry}

\begin{abstract}
The ultrafast photochemistry of heterocyclic compounds is central to photobiology and materials chemistry, yet direct experimental observation of their structural dynamics remains rare. Here we present the first real-time structural characterization of photoinduced ring opening and subsequent fragmentation in the isomeric pair oxazole and isoxazole using MeV ultrafast electron diffraction (UED), complemented by non-adiabatic molecular dynamics simulations. 
Upon photoexcitation at \SI{200}{\nano\meter}, both isomers undergo ring opening followed by fragmentation into various products, but with strikingly different dynamics governed by heteroatom positioning. Trajectory Surface Hopping (TSH) simulations at the SA(4)-CASSCF(12,10)/aug-cc-pVDZ level reproduce the experimental diffraction signatures, which are surprisingly similar for both isomers, and reveal distinct mechanistic pathways for the two isomers. For isoxazole, all trajectories exclusively undergo N--O bond cleavage within ${\approx}40$~fs, followed by sequential fragmentation into HCN~+~ketene and HCO~+~vinyl nitrene channels on the hundreds-of-femtoseconds timescale. Oxazole, by contrast, shows significantly slower ring opening (${\approx}290$~fs) with only 85\% efficiency, proceeding primarily through O--C cleavage and accessing a richer landscape of intermediates including nitrile ylide and O-pyramidalized structures. Simulated diffraction patterns derived from trajectory ensembles, convolved with the experimental instrument response function, are in agreement with the branching ratios observed by UED. Classification of conical intersection topographies reveals that isoxazole funnels to the ground state through vinyl nitrene and open-chain geometries at energetically accessible crossings, whereas oxazole navigates through nitrile ylide intermediates and puckered ring structures. This synergy between UED and trajectory surface hopping provides an atomistic picture of how the simple interchange of heteroatom connectivity in structural isomers fundamentally reshapes excited-state potential surfaces, conical intersection accessibility, and photochemical outcome.
\end{abstract}

\section{Introduction}

Five-membered heterocyclic compounds containing nitrogen and oxygen atoms are ubiquitous in biological systems, pharmaceuticals, and functional materials, where their photochemical properties play crucial roles in processes ranging from DNA photodamage and repair to organic photovoltaic device operation.\cite{Kwok2006,Zhang2015,Turro2010,GomezBombarelli2016,Schreier2007} These heterocycles often undergo rapid ring-opening reactions upon UV excitation, leading to reactive intermediates that can participate in subsequent photochemical transformations. Understanding the fundamental mechanisms governing these ultrafast processes is essential for controlling photochemical outcomes and designing molecules with tailored photophysical properties.

\begin{figure}[htbp]
\centering
\includegraphics[width=\textwidth]{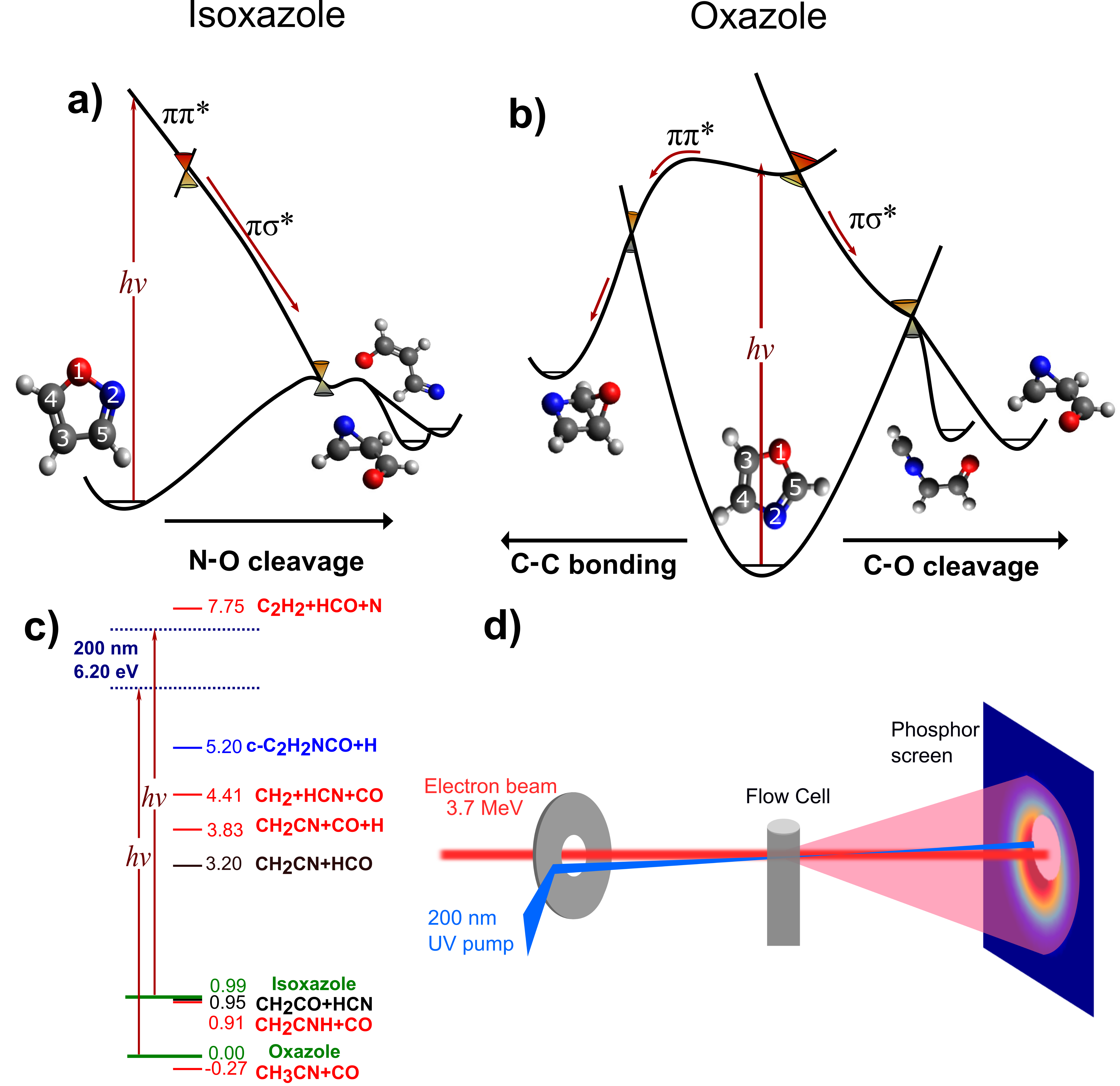}
\caption{Schematic potential energy diagrams for (a) isoxazole and (b) oxazole. The vertical red arrow represents the \SI{200}{\nano\meter} excitation onto the \textpi\textpi* state. (c) Asymptotic product energies relative to oxazole (taken from Ref.~\cite{dais_2018_iso} and Ref.~\cite{ATcT_website}) of the fragmentation pathways observed in our theoretical calculation. Green color indicates the starting molecules, red indicates fragments only observed in the isoxazole calculation, blue indicates fragments only observed in the oxazole calculations, and black indicates fragments observed in both calculations. (d) Schematic of the experimental setup showing the \SI{200}{\nano\meter} pump and MeV-electron probe beams overlapped nearly co-linearly inside the gas sample prepared in a flow cell. The diffracted electrons are detected on a phosphor screen detector (see Methods for details).}
\label{fig:scheme}
\end{figure}

Oxazole and isoxazole (\ce{c-C3H3NO}) are structural isomers -- five-membered rings differing only in the relative positions of the nitrogen and oxygen atoms (see Figure~\ref{fig:scheme}a and b). With readily accessible UV excitations and a rich array of product channels, they represent a prototypical isomer pair to reveal the impact of structural differences on photodynamical pathways. The global energy minimum is oxazole, in which the nitrogen and oxygen atoms are separated by a carbon atom, while isoxazole, with the heteroatoms adjacent, lies approximately \SI{1}{\electronvolt} higher in energy. In both molecules, excitation by a \SI{200}{\nano\meter} photon promotes an electron to the \textpi\textpi* and n\textpi* states, with an additional Rydberg state accessible in oxazole. From there, the structural differences give rise to profound differences in the electronic character and decay pathways as shown schematically in Figure~\ref{fig:scheme}.

Previous theoretical work has explored the ultrafast dynamics of both molecules following \textpi\textpi* excitation. Using nonadiabatic direct dynamics calculations, Cao \textit{et al.}\cite{Cao_2015_isoxazole, CAO_2016_oxazole} found that isoxazole exclusively showed ring opening at the N--O bond. This ring opening led either to rapid fragmentation or to azirine ring formation, with both pathways controlled by a low-lying intermediate structure near the S$_1$/S$_0$ conical intersection (CI). Oxazole showed more complex dynamics: most trajectories (81\%) underwent ring opening at the O$_1$--C$_5$ bond, while a minor pathway (16\%) involved C$_3$--C$_5$ bond formation giving either a bicyclic intermediate or relaxation back to oxazole forming a hot ground-state molecule. Geng \textit{et al.}\cite{geng_2020} also performed trajectory calculations to complement their time-resolved photoelectron spectroscopy (TRPES) measurements. Their simulation results were in overall agreement with the previous calculations by Cao \textit{et al.} but showed a lower fraction of ring opening in oxazole and a faster time scale, which was biased by the fact that they only simulated the first \SI{100}{\femto\second}.  

Given the large excess energy following UV excitation, it is inevitable that fragmentation will follow ring opening. From final product branching measurements,\cite{dais_2018_iso,DownesWard2024} it is known that oxazole and isoxazole give the same primary products, HCN + \ce{CH2CO} (70\% and 53\% branching for oxazole and isoxazole, respectively); however, their secondary products differ. In oxazole, the secondary pathway (23\%) is HCO + \ce{CH2CN}/\ce{CH2NC}; while for isoxazole, it is CO + \ce{CH3CN} (23\%). Based on comparison to Cao's work and pyrolysis studies augmented by extensive characterization of the ground-state surface,\cite{Nunes2011} the secondary pathway was hypothesized to be slower, resulting from more extended exploration of the ring-open region of the potential surface before fragmenting. Both pathways in isoxazole are thought to be controlled by a low-lying ring-open minimum near the S$_1$/S$_0$ conical intersection. No equivalent structure is present on the oxazole potential energy surface, accounting for the differing secondary pathways.

While these prior studies have provided valuable insights into the early-time electronic relaxation, several several gaps in our understanding remain. TRPES measurements, though sensitive to electronic structure changes, provide only indirect information about molecular geometry. The trajectory simulations were limited to \SI{100}{\femto\second}, precluding observation of subsequent fragmentation dynamics. Furthermore, no experimental technique has directly captured the structural evolution accompanying ring opening and bond-breaking in real time.

In this work, we address these gaps by presenting the first direct structural observation of the ultrafast photodynamics of oxazole and isoxazole using MeV electron diffraction (UED). 
We complement these measurements with extended nonadiabatic molecular dynamics simulations propagated to \SI{1}{\pico\second}, revealing the complete mechanistic picture from initial excitation through final product formation. 
Using simulated diffraction patterns generated from trajectory ensembles to fit the experimental data, we are able to extract time-dependent structural populations from the UED data and identify ring opening, intermediate formation, and fragmentation.
This combined approach reveals significant differences in the photodynamical behavior of the two isomers, with isoxazole showing ultrafast N--O bond cleavage, which occurs within approximately \SI{40}{\femto\second} according to our calculations, while oxazole undergoes slower ring opening with an average time of \SI{290}{\femto\second} and more complex structural changes. We show that the heteroatom position controls the dynamics through its impact both on the initial electronic decay pathways and on the available structures accessed en route to ring opening and fragmentation.

\section*{Results and Discussion}

In this and the following Sections, we first present the experimental MeV-UED diffraction signals and compare them to the simulated diffraction signal obtained from theory. We then discuss the results of the theoretical trajectories for each isomer in more detail, including characteristic intermediates, ring-opened structures, and fragments. These are then used to construct simulated basis diffraction patterns from which time-dependent populations are extracted by fitting to the experimental data. 

\begin{figure}[t!]
\centering
  \includegraphics[width=1.0\textwidth]{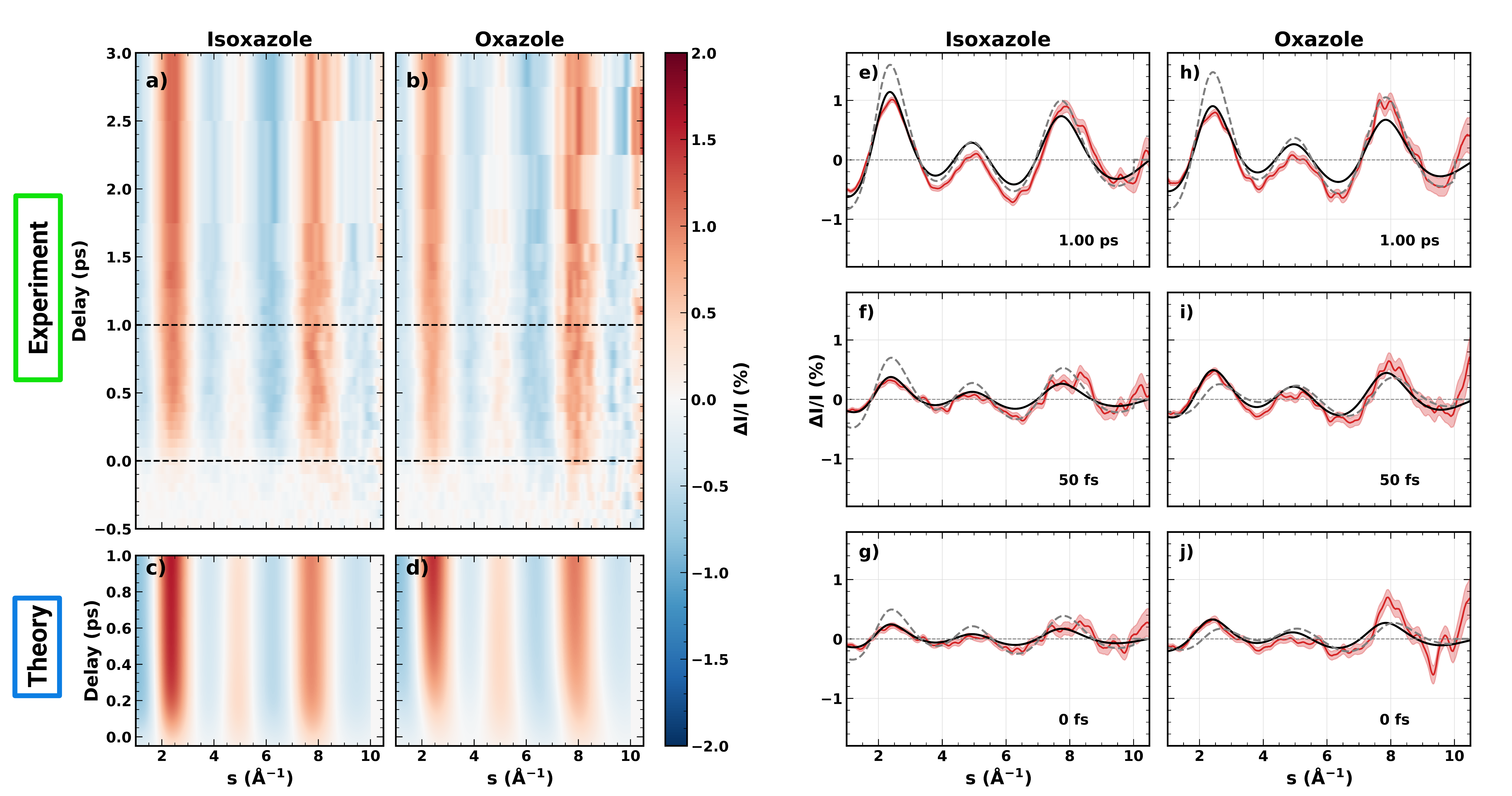}
\caption{Time-dependent difference-diffraction intensity $\Delta I/I$ (\%). (a) and (b) are the experimental results for isoxazole and oxazole, respectively. (c) and (d) are the corresponding theoretical simulations for isoxazole and oxazole. The dashed horizontal lines in (a) and (b) indicate the temporal range covered by the theoretical simulations. All plots share the same color scale. Plots on the right are lineouts at the indicated time delays (\SI{1}{\pico\second}, \SI{50}{\femto\second}, and \SI{0}{\femto\second}), with experimental data in red, basis function fits in black, and theoretical simulations as gray dashed lines. Basis function fitting was performed over the momentum transfer range $s = 1.7$--$8.6$~\AA$^{-1}$, excluding the low-$s$ region affected by the beam stop and the high-$s$ region dominated by noise. Plots for isoxazole are shown in (e)--(g), while those for oxazole in (h)--(j). The red shaded region represents $\pm 1\sigma$ uncertainty of the experimental data estimated via bootstrapping over 150 experimental replicates. The theoretical curves are scaled by the fitted excitation fraction.}
\label{fig:LC_data_sim}
 \end{figure}
The MeV-UED experiment was conducted in two modes: high electron bunch charge with longer electron bunches giving better signal to noise but lower time resolution, and low electron bunch charge, affording improved temporal resolution but lower signal. Experimental details are given in the Methods section and in Section~1 of the Supplementary Information (SI). All of the data in the Results and Discussion section and Section~2 of the SI were recorded in low-charge mode, while high-charge results are shown in Section~3 of the SI. Figure~2 shows the experimental difference-diffraction signal, $\Delta I/I$, obtained from data for isoxazole (a) and oxazole (b) using the analysis procedure described in Section~1 of the SI. The red areas in these difference plots indicate regions where an increased scattering intensity was observed, while blue shows a decrease in intensity. We observe a maximum change in intensity of about $\approx$1~\% of the static signal. Overall, the experimental data for the two isomers are remarkably similar, given the difference in expected dynamics and absorption cross-section. In particular, there are several key features observed for both isomers. Both show a small increase in intensity at around 0.6~\AA$^{-1}$~ at $t_0$ (see Supplementary Figure~S1), which is attributed to dynamics of the valence electron upon excitation in accordance to similar observations in previous MeV-UED measurements \cite{yang2020simultaneous}. Attosecond electronic dynamics in oxazole has also recently been the focus of several studies \cite{mukamel2021,giri2023attosecond}. As this feature in our data is the result of the breakdown of the independent-atom model (IAM), it is not reproduced by the theoretical simulation of the UED signal, which relies on the IAM. Another interesting time-dependent feature is a slight narrowing of the red region at 5 \AA$^{-1}$, which is visible in the experiment and simulations for both isomers. The only clear difference in the diffraction plots between the two isomers is the red feature at 2.5~\AA$^{-1}$, which increases more quickly for isoxazole as compared to oxazole in both experiment and simulation (see lineouts in panels (c)-(f) in Supplementary Figure~S1). 

Simulated time-dependent diffraction difference signals are shown in Figure \ref{fig:LC_data_sim}(c),(d). Corresponding $\Delta I/I$ snapshots at three different time delays for both isomers for experiment (red lines) and theoretical simulations (gray dashed lines) are shown in Figure \ref{fig:LC_data_sim}(e)-(j). The black lines show the results of a least-square fit to the experimental data, which is explained in more detail in the Section ``Extracting time-dependent structural dynamics from the UED data''. Overall, the theoretical simulations match the experimental results well, with peaks and troughs at the same scattering angles as well as qualitatively similar temporal patterns. The magnitude of the difference-diffraction signal in oxazole is slightly smaller and the time scale of the changes appears to be slower than in isoxazole, most obviously for the 2.5 and 8~\AA$^{-1}$ features. While the isoxazole and oxazole experimental data are otherwise almost identical, the theoretical simulations reveal dramatic underlying differences in both the dynamics and the mechanisms responsible.


\subsection{Trajectory pathways}

To connect the observed diffraction signatures to the underlying photochemistry, we analyze the trajectory simulations and identify the distinct reaction pathways governing ring opening and fragmentation in each isomer (see Sections~4-8 of the SI for details). We find that the two structural isomers undergo dramatically different dynamics that can be attributed to the electronic character of the initially populated excited states, the topography in the vicinity of the conical intersections mediating nonadiabatic relaxation, and the subsequent bonding opportunities available in the potential products.

\begin{figure}[htbp]
\centering
\includegraphics[width=\textwidth]{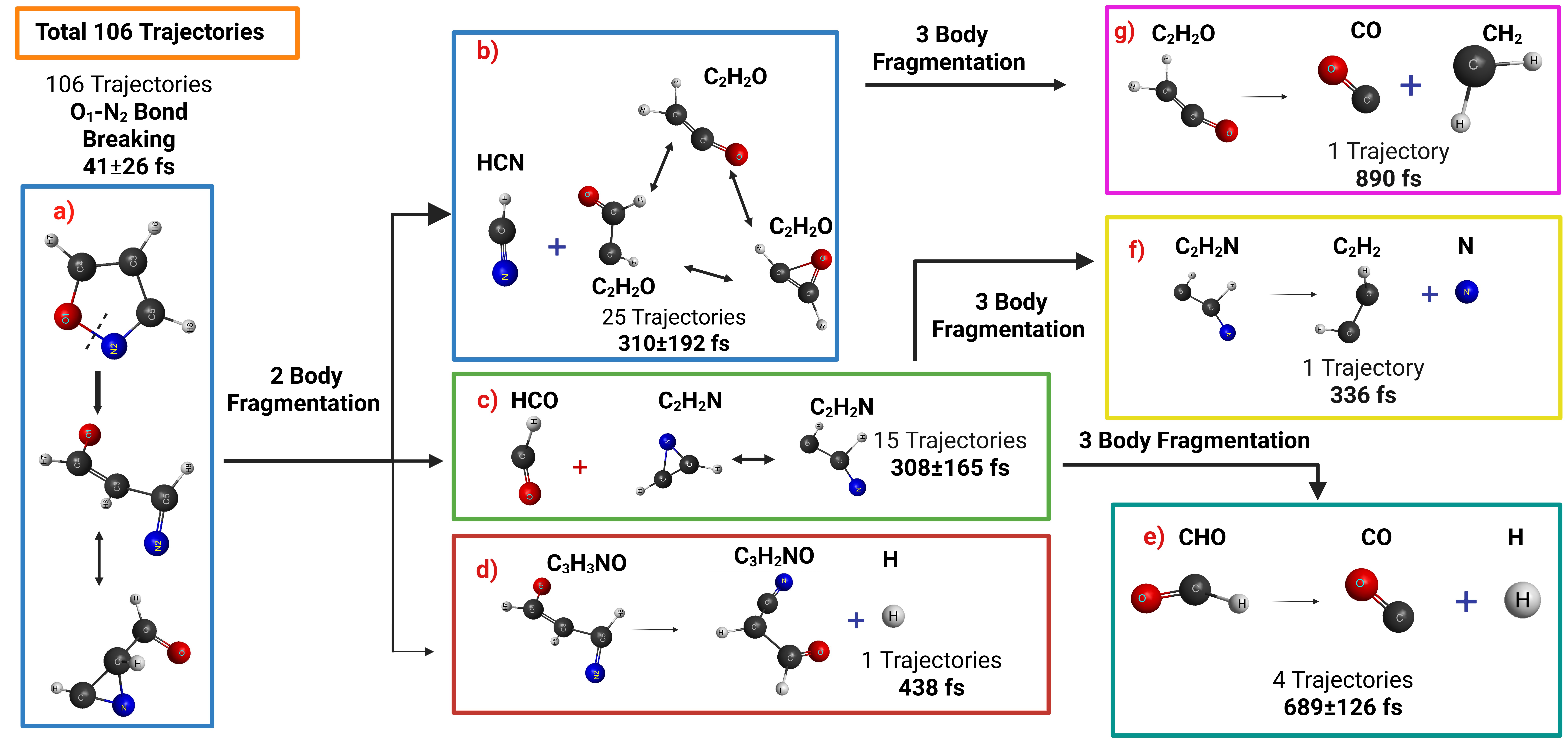}
\caption{Schematic representation of the dominant fragmentation pathways observed in the nonadiabatic dynamics of isoxazole following photoexcitation (106 trajectories, up to \SI{1000}{\femto\second}). (a) The primary event involves N--O bond cleavage, observed in 106 trajectories (100\%) with an average onset time of $41 \pm 26$~fs, leading to ring-opened intermediates and transient azirine formation. (b) Formation of HCN + \ce{C$_2$H$_2$O} fragments occurs in 25 trajectories (24\%) with an average time of $310 \pm 192$~fs; 1 trajectory proceeds to three-body fragmentation. (c) The HCO + \ce{C$_2$H$_2$N} channel represents a second two-body pathway, observed in 15 trajectories (14\%) with an average time of $308 \pm 165$~fs; 5 trajectories proceed to three-body fragmentation. (d) H loss pathway (1 trajectory, 1\%, \SI{438}{\femto\second}). (e) Three-body fragmentation producing CHO + CO + H (4 trajectories, 4\%, $689 \pm 126$~fs). (f) Three-body fragmentation producing \ce{C$_2$H$_2$} + N (1 trajectory, 1\%, \SI{336}{\femto\second}). (g) Three-body fragmentation producing CO + \ce{CH$_2$} from ketene dissociation (1 trajectory, 1\%, \SI{890}{\femto\second}). At \SI{1}{\pico\second}, 65 trajectories (61\%) remain as ring-opened structures without fragmentation.}
\label{fig:isoxazole}
\end{figure}

Figure~\ref{fig:isoxazole} illustrates the photodynamical pathways for isoxazole derived from 106 non-adiabatic trajectories propagated for up to \SI{1000}{\femto\second}. Upon UV excitation, the majority of trajectories are initiated on the S$_1$ surface, which possesses dominant $\pi\pi^*$ character at the Franck--Condon geometry (Tables~S2 and~S4). A smaller fraction of trajectories begins on S$_2$, which carries $n\pi^*$ character (Figure~S16). This S$_1$($\pi\pi^*$)/S$_2$($n\pi^*$) assignment is consistent with the earlier studies of
Cao\cite{Cao_2015_isoxazole} and Geng \textit{et al.}\cite{geng_2020}, both of which identified
a bright $\pi\pi^*$ state as S$_1$ and an $n\pi^*$ state as S$_2$ for isoxazole. Although neither state is directly dissociative at the Franck--Condon geometry, the adjacent placement of nitrogen and oxygen creates a weak N--O bond along which a low-lying $\pi\sigma^*$ state becomes rapidly accessible upon slight nuclear displacement.
Upon excitation, isoxazole thus undergoes ultrafast transfer to a dissociative $\pi\sigma^*$ state localized along the N--O bond, in agreement with previous theoretical and experimental studies.\cite{Cao_2015_isoxazole,geng_2020} Consistent with this electronic structure picture, N--) bond cleavage occurs in all trajectories within approximately 40~fs on average, establishing this as the exclusive ring-opening pathway. The classification of molecular geometries at surface hopping points (see Table~S7 and Figure~S36) provides direct insight into the conical intersection topography mediating this process. At the S$_2$/S$_1$ crossing, the presence of ring-intact structures with early N--O bond elongation and vinyl nitrene character indicates that the S$_2$ to S$_1$ transition occurs along a coordinate that is continuous with the ring-opening motion. At the S$_1$/S$_0$ crossing, the vinyl nitrene geometry dominates, confirming that passage through the S$_1$/S$_0$ conical intersection is dominated by prior ring opening and formation of the open-chain nitrene intermediate. Additional S$_1$/S$_0$ hops occur at already-fragmented HCN + \ce{C2H2O} geometries, indicating that some trajectories undergo bond dissociation on the excited-state surface before reaching the ground state.

Following ring opening, a fraction of the trajectories form a transient azirine intermediate before fragmentation. Subsequent fragmentation proceeds through two dominant two-body channels with comparable onset times of approximately 310~fs: HCN + \ce{C$_2$H$_2$O} (24\%) and HCO + \ce{C$_2$H$_2$N} (14\%). Minor pathways include H loss and three-body fragmentation ($<$6\% combined), the latter occurring via secondary dissociation on longer timescales (see Figure~\ref{fig:isoxazole} for details). At \SI{1}{\pico\second}, roughly half the trajectories remain as ring-opened structures that have not yet fragmented, with the remainder distributed among two-body fragments (33\%), azirine intermediates (10\%), and three-body products (6\%).

Prior experimental product branching measurements \cite{DownesWard2024} indicate that HCN + \ce{C$_2$H$_2$O} is the dominant fragment channel, consistent with our simulations. However, a significant experimental contribution (23\%) from \ce{CH$_3$CN} + CO \cite{DownesWard2024} is absent in our calculations. This pathway requires extensive hydrogen migration and C--C bond rearrangement mediated by a vinylnitrene intermediate,\cite{Nunes2011} processes whose timescales typically exceed several picoseconds due to the need for intramolecular vibrational energy redistribution (IVR) into the relevant reaction coordinates,\cite{Nesbitt1996,Keshavamurthy2020} and which therefore fall outside the \SI{1}{\pico\second} window covered by our calculations. This interpretation is supported by comparisons between photolysis and pyrolysis product distributions, where the latter shows enhanced \ce{CH$_3$CN} + CO yields due to complete statistical energy redistribution.\cite{DownesWard2024,Nunes2011}

\begin{figure}[htbp]
\centering
\includegraphics[width=\textwidth]{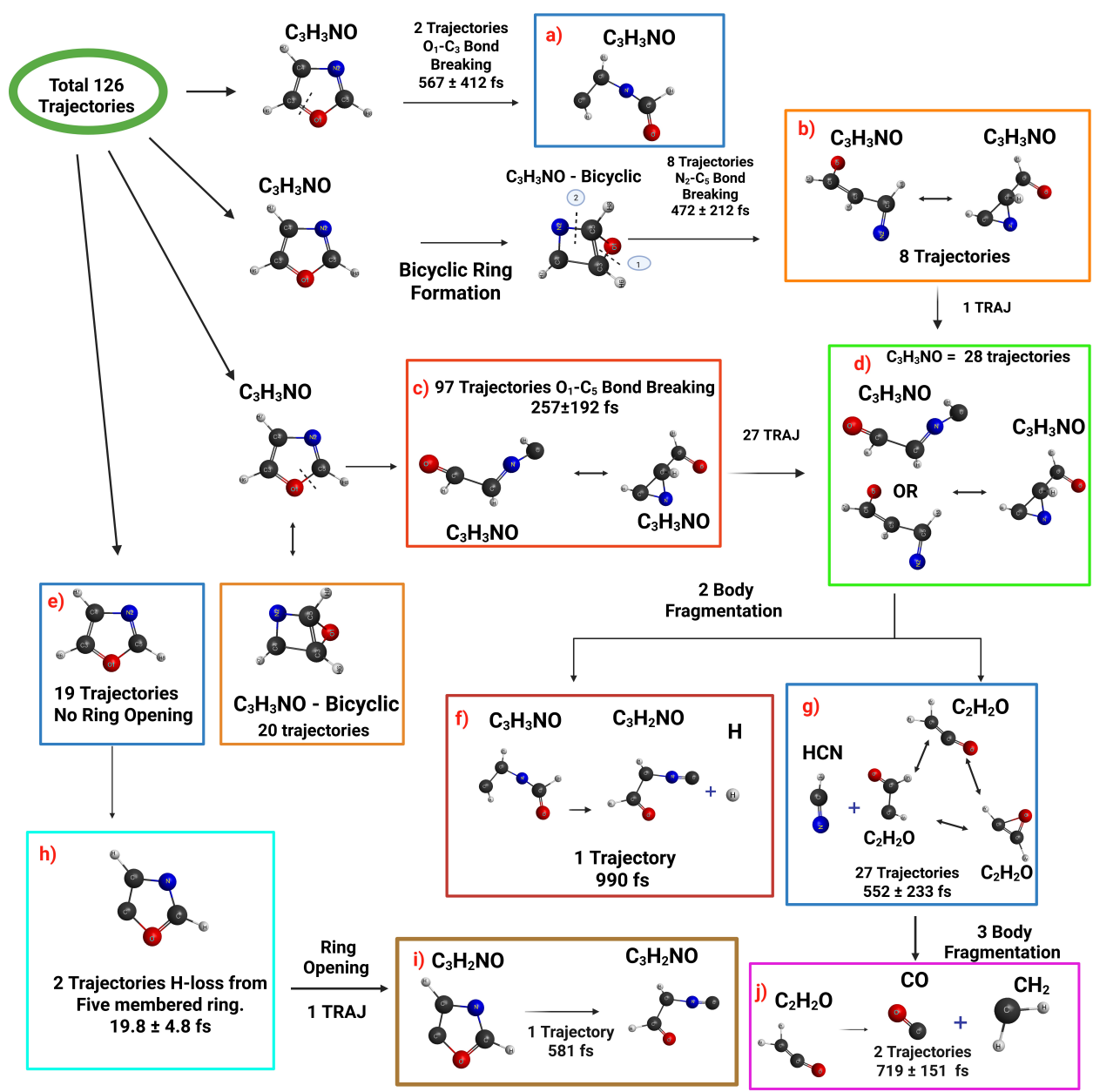}
\caption{Schematic representation of the photodynamical pathways observed in the nonadiabatic dynamics of oxazole following photoexcitation (126 trajectories, up to \SI{1000}{\femto\second}). Ring opening occurs in 108 trajectories (85\%) through multiple channels: (c) O$_1$--C$_5$ bond cleavage (97 trajectories, 77\%, $257 \pm 192$~fs), (a) O$_1$--C$_3$ bond cleavage (2 trajectories, 2\%, $567 \pm 412$~fs), and (b) N$_2$--C$_5$ bond breaking via bicyclic intermediate (8 trajectories, 6\%, $472 \pm 212$~fs). (e) 19 trajectories (15\%) do not undergo ring opening. Bicyclic ring formation occurs in 28 trajectories total. Ring-opened structures lead to (d) an open-chain intermediate (28 trajectories), which can undergo (g) two-body fragmentation to HCN + \ce{C2H2O} (27 trajectories, $552 \pm 233$~fs) or (f) H loss (1 trajectory, 990~fs). (h) H loss from the five-membered ring (2 trajectories, $19.8 \pm 4.8$~fs) can lead to (i) ring opening (1 trajectory, 581~fs). (j) Three-body fragmentation producing CO + \ce{CH2} + HCN (2 trajectories, $719 \pm 151$~fs) occurs via ketene dissociation.}
\label{fig:oxazole}
\end{figure}
Oxazole exhibits markedly different and more complex dynamics compared to its structural isomer, a contrast that originates in the fundamentally different electronic structure of its low-lying excited states. Figure~\ref{fig:oxazole} summarizes the photodynamical pathways derived from 126 nonadiabatic trajectories. In contrast to isoxazole, where the bright state has intrinsic dissociative character, the low-lying excited states of oxazole lack a direct $\pi\sigma^*$ channel along any ring bond. At the Franck--Condon geometry (Table~S3), the S$_1$ state has Rydberg character, with the Natural Difference Orbital  (NDO) attachment density extending into diffuse space rather than localizing along a specific bond (Figure~S24), while S$_2$ is an $n\pi^*$ state (Figure~S25) and S$_3$ carries $\pi\pi^*$ character (Figure~S26). The absence of a direct dissociative channel from the Franck–Condon region stands in stark contrast to isoxazole, where the πσ* state is immediately accessible along the weak N–O bond. In oxazole, accessing the ring-opening pathway requires the molecule to first undergo non-adiabatic population transfer and substantial nuclear rearrangement before accessing dissociative regions of the potential energy surface. Cao\cite{CAO_2016_oxazole} placed an $n_\mathrm{N}\pi^*$ state lowest at the MS-CASPT2 level, with $\pi\pi^*$ states as S$_2$ and S$_3$, whereas Geng \textit{et al.}\cite{geng_2020} found a bright $\pi\pi^*$ state as S$_1$, an $n\pi^*$ state as S$_2$, and a $\pi\sigma^*$ state as S$_3$ at the ADC(2) level. Our calculations instead assign a diffuse Rydberg state as S$_1$, with $n\pi^*$ (S$_2$) and $\pi\pi^*$ (S$_3$) above it. The emergence of this low-lying Rydberg state is enabled by the diffuse functions of the aug-cc-pVDZ basis employed here; Geng \textit{et al.} have also reported a dense manifold of Rydberg states near the Franck--Condon geometry (within which the reactive $\pi\sigma^*$ state is buried). Despite these differences in  ordering, all three studies agree that no directly dissociative $\pi\sigma^*$ channel is available from the Franck--Condon region of oxazole, consistent with the slower, multi-step ring opening observed here.

The conical intersection geometries accessed during the dynamics (see Table~S7 and Figure~S37) reveal the mechanistic pathway through which oxazole navigates this more complex landscape. At the S$_3$/S$_2$ crossing, the vast majority retain an intact ring, with a small fraction showing O-atom pyramidalization, an out-of-plane distortion of the oxygen atom from the ring plane that facilitates nonadiabatic transitions. This pyramidalization becomes increasingly prominent at lower-energy crossings: at the S$_2$/S$_1$ intersection, 16\% of hop geometries exhibit O-pyramidalization alongside nitrile ylide (ring-opened) character. By the S$_1$/S$_0$ crossing, the nitrile ylide geometry dominates, with O-pyramidalized ring-intact structures comprising an additional 20\%. This progression from planar ring-intact geometries at the highest crossing, through O-pyramidalized structures at intermediate crossings, to ring-opened nitrile ylide intermediates at the S$_1$/S$_0$ funnel maps out the sequential nuclear distortions required for oxazole to reach the ground state. Consistent with this electronic structure analysis, the ring-opening process is considerably slower than in isoxazole, with an average onset time of approximately 290~fs in oxazole compared to 40~fs in isoxazole. Of the 126 simulated oxazole trajectories, 85\% undergo ring opening, predominantly through O$_1$--C$_5$ bond cleavage (77\% of all trajectories; $257 \pm 192$~fs), with minor contributions from O$_1$--C$_3$ cleavage (2\%) and N$_2$--C$_5$ bond scission via a bicyclic intermediate (6\%; $472 \pm 212$~fs). Atom numbering follows the labeling scheme defined in Figure~1. A notable feature of the oxazole photodynamics is the formation of bicyclic intermediates by C$_3$--C$_5$ bond formation across the ring, generating a strained three-membered ring, as previously reported by Cao.\cite{CAO_2016_oxazole} Formation of the bicyclic intermediate is reversible: of the 28 trajectories that visit this geometry, 8 subsequently ring-open via N$_2$-- C$_5$ bond scission, while the remainder equilibrate between the closed ring and bicyclic ring. A further 14\% of the trajectories do not undergo ring opening within \SI{1}{\pico\second}---a direct consequence of the absence of a barrierless dissociative channel from the Franck--Condon region. Two trajectories lose a hydrogen atom from the intact ring on a remarkably fast timescale ($19.8 \pm 4.8$~fs), with one subsequently undergoing ring opening at 581~fs. Two-body fragmentation into HCN + \ce{CH2CO} occurs in 21\% of trajectories at an average onset of $552 \pm 233$~fs. Minor channels include H-atom elimination (three trajectories in total) and three-body fragmentation (2\%), the latter producing CO + \ce{CH2} + HCN via secondary dissociation of the ketene fragment ($719 \pm 151$~fs).

At \SI{1}{\pico\second}, the population distribution reflects the slower, more complex dynamics of oxazole, with significant proportion persisting as ring-opened structures and two-body fragments, with the remainder distributed among bicyclic intermediates, intact rings, and minor channels. The persistence of multiple intermediate structures throughout the simulation is a consequence of the more complex excited-state potential energy landscape of oxazole, where competing relaxation pathways and accessible conical intersection seams distribute population across several long-lived structural motifs rather than funneling it through a single dominant channel.

\subsection{Extracting time-dependent structural dynamics from the UED data}
\begin{figure}[htbp]
\centering
\includegraphics[width=\textwidth]{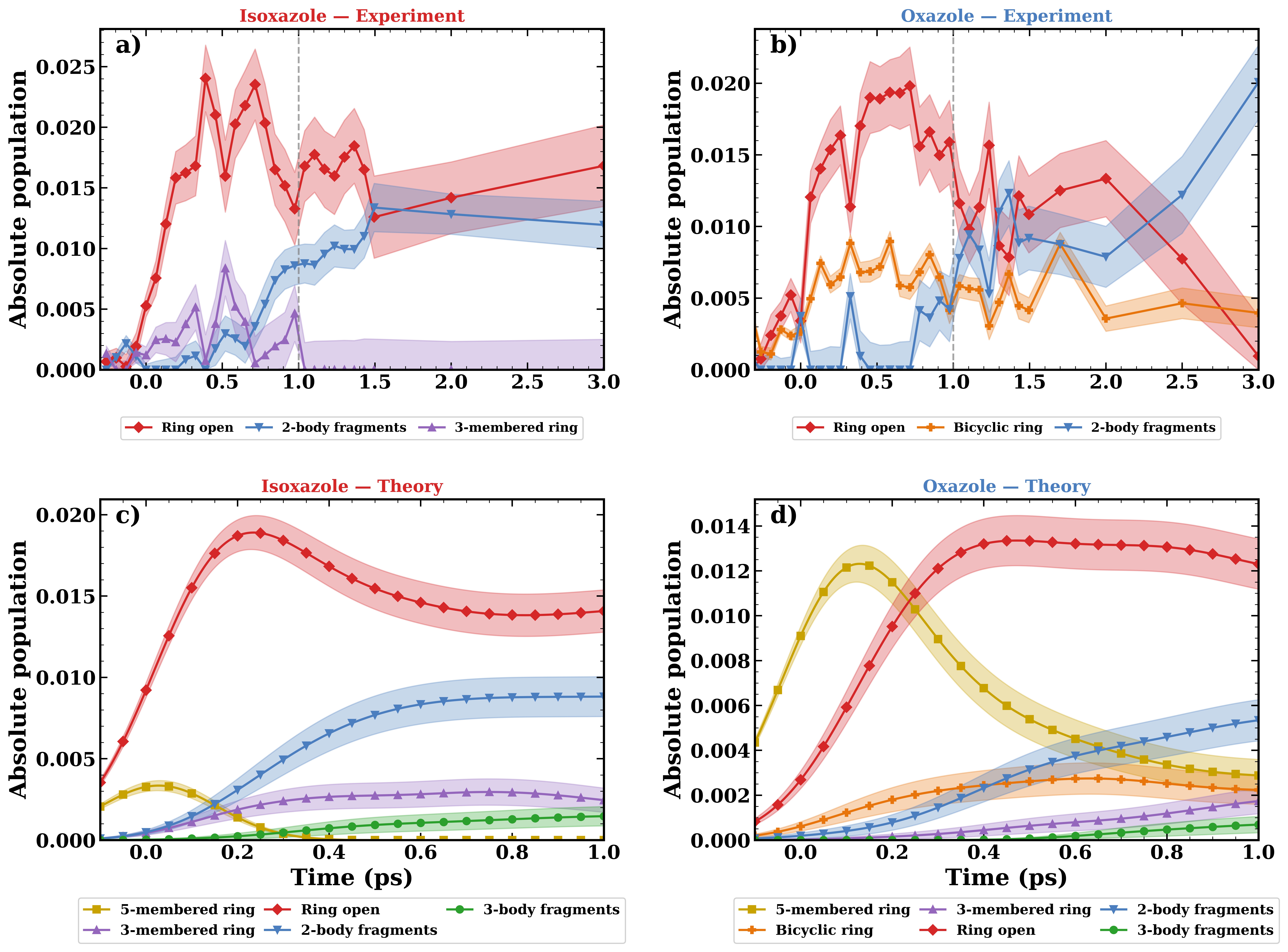}
\caption{
Time-dependent absolute population dynamics from UED basis function fitting (a, b) and convolved trajectory simulations (c, d). Note the different delay ranges for experiment (up to \SI{3}{\pico\second}) and simulations (up to \SI{1}{\pico\second}). The vertical dashed gray line in panels (a, b) marks the \SI{1}{\pico\second} limit of the theoretical results shown in panels (c, d). Absolute populations are obtained by multiplying the fitted structural fractions by the excitation fraction at each delay. (a)~Isoxazole experimental populations showing rapid ring opening (red) as the dominant channel, followed by gradual growth of 2-body fragments (blue). The 3-membered ring (azirine, purple) shows a transient signal at early delays ($<$\SI{1}{\pico\second}). (b)~Oxazole experimental populations revealing more complex multi-pathway dynamics: ring opening (red) dominates the first ${\approx}$\SI{2}{\pico\second} but with larger fluctuations than isoxazole, the bicyclic ring (orange) maintains a persistent population throughout the \SI{3}{\pico\second} window, and 2-body fragments (blue) grow steadily after ${\approx}$\SI{0.8}{\pico\second}, overtaking the ring-open population at ${\approx}$\SI{2.4}{\pico\second}. (c)~Convolved isoxazole theory populations scaled by the experimental excitation fraction (${\approx}2.7\%$). The 5-membered ring parent (golden) depletes within ${\approx}$\SI{300}{\femto\second}, ring opening (red) peaks at ${\approx}$\SI{250}{\femto\second} and remains the dominant population throughout, 2-body fragments (blue) grow to ${\approx}0.009$, and the 3-membered ring (purple) stabilizes at ${\approx}0.003$. The 3-body fragments (green) remain minor throughout. (d)~Convolved oxazole theory populations, scaled by the experimental excitation fraction (${\approx}2.5\%$), showing slower dynamics: the 5-membered ring (golden) reaches its maximum at ${\approx}$\SI{150}{\femto\second} and depletes steadily thereafter, ring opening (red) rises to ${\approx}0.013$ by ${\approx}$\SI{450}{\femto\second}, while the 2-body fragments (blue), bicyclic ring (orange), and 3-membered ring (purple) grow gradually through \SI{1}{\pico\second}. Theory populations extend below \SI{0}{\pico\second} due to the convolution with the \SI{300}{\femto\second} instrument response function.}
\label{fig:branching}
\end{figure}

To extract quantitative information about the time-dependent yields of various intermediate and product geometries, we employed a least-square fitting procedure using the geometry information from the trajectories, similar to the procedure developed recently by Nunes \textit{et al.}\cite{nunes2023monitoring}.
First, the geometries from the trajectory calculations were categorized into different groups using a random forest machine learning algorithm (further details are given in Section~9A of the SI). For isoxazole, these groups were the 5-membered ring, the ring-open structure, the 3-membered azirine ring, and several fragment pathways shown in Fig.~\ref{fig:isoxazole}. The average UED scattering pattern for each group was simulated using the IAM (see Figure~S3). For a given fragmentation channel, the UED scattering signal was calculated as the (incoherent) sum of the scattering patterns of the individual fragments. Since UED is less sensitive to hydrogen atoms and unable to distinguish between the C, N, and O atoms, the fragment pathways were grouped by the number of heavy atoms in each fragment. This partitioned the isoxazole fragments into two groups: A two-body fragmentation group comprising one fragment with 3 heavy atoms and the other with 2 heavy atoms, labeled 2+3, and a three-body fragmentation group comprising two fragments with 2 heavy atoms each and one fragment with one heavy atom, labeled 2+2+1. The UED patterns for all pathways within their respective groups were very similar, but the group patterns were distinct. For each group, a representative scattering pattern was calculated by taking an average over each pathway's time-averaged UED pattern (see Figure~S3a).
These representative scattering patterns for the different geometry groups, summarized in Figure~S38, were then used as basis functions to fit the experimental data between 1.70 and 8.58~\AA$^{-1}$. The 5-membered ring scattering pattern was excluded from the fit due to the lower amplitude of the scattering pattern. The amplitude for each function was varied to fit the experimentally measured scattering pattern at each delay point, giving the fraction of each structure at a given delay, as shown in Fig.~\ref{fig:branching}a. 
The interpretation of these results is discussed in more detail below together with the fit results for oxazole. The fits also provide an estimate of the overall excitation fraction, which was 2.7\% and 2.3\% for isoxazole and oxazole, respectively. 

As discussed above, oxazole showed a bicyclic ring in the trajectories, thus necessarily requiring an additional basis function in the fits. However, given that the UED pattern for the 3-membered azirine ring is very similar to that of the ring-open structure (see Figure~S3), we found that a fit with four basis functions became unstable for oxazole, so we had to remove the 3-membered azirine ring structure, which is a minority product according to our trajectory simulations, from the fit in order to obtain stable fit results. Furthermore, for oxazole, fragmentation proceeded almost exclusively through the 2+3 channel: two-body fragmentation into HCN + \ce{CH2CO} occurred in 27 of the 126 trajectories, while three-body fragmentation (CO + \ce{CH2} + HCN, via secondary dissociation of the ketene fragment) was observed in only two trajectories, and H elimination in three. No trajectories produced the alternative HCO + \ce{CH2CN} channel, consistent with prior experimental observations identifying HCN + ketene as the dominant final-product channel \cite{DownesWard2024}. Therefore, we used the HCN + \ce{CH2CO} scattering pattern for the fits to the 2+3 fragment pathway.  


Figure \ref{fig:branching} shows a comparison of the absolute population of each geometry for isoxazole (a,c) and oxazole (b,d) obtained from both the above fit procedure and from the trajectory simulations. 
Note that the experimental data in panels (a) and (b) show the fits out to 3~ps, while the data from the simulations in panels (c) and (d) extends only to 1~ps. For isoxazole, the fit to the experimental data shows a very rapid ring opening, followed by a slow decrease in the ring-open population while the population of 2+3 fragments increases correspondingly. At a delay of ${\approx}1.5$~ps, the ring-open and fragment populations become equal, each accounting for about half of the excited molecules. The 3-membered azirine ring contributes only transiently at early delays ($<$\SI{1}{\pico\second}) before decaying to zero. Figure \ref{fig:branching}(c) shows the corresponding theoretical populations for isoxazole, convolved with a 300~fs Gaussian to match the experiment. The theory shows the same sequence: rapid ring opening followed by fragmentation. The onset of fragmentation, however, occurs considerably earlier in the simulations, with the 2+3 population growing within ${\approx}200$~fs of excitation and plateauing at roughly one third of the excited molecules by ${\approx}0.7$~ps, whereas the experimental fragment signal only rises after ${\approx}0.5$~ps. Within the 1~ps simulation window, the ring-open structure remains the dominant population, consistent with the 61\% of trajectories that have not yet fragmented at 1~ps (Figure~\ref{fig:isoxazole}). 

Figure \ref{fig:branching}(b) shows the experimental populations for oxazole. As explained above, the fits for oxazole require a different basis, as structures were observed in the trajectory calculations that cannot occur for isoxazole. The oxazole basis includes the bicyclic ring, formed by O-pyramidalization followed by C--C bond formation. Its fitted population rises within the first few hundred femtoseconds to roughly 30\% of the excited molecules and
persists throughout the full 3~ps measurement window. The ring-open structure dominates the first
${\approx}2$~ps before declining, while the HCN+\ce{CH2CO} (2+3) fragments grow steadily after ${\approx}0.8$~ps, overtaking the ring-open population at ${\approx}2.4$~ps. As mentioned above, the 3-membered azirine ring was not included in the final oxazole basis: its scattering pattern is too strongly correlated with that of the ring-open structure, and substituting it for the bicyclic basis yields comparable fit quality but unstable, unphysical amplitudes. Figure \ref{fig:branching}(d) shows the theoretical results. The five-membered ring depletes mainly into the ring-open structure, which rises until ${\approx}450$~fs and then plateaus, while the 2+3 fragments grow steadily throughout the simulated window. The bicyclic and 3-membered ring structures contribute only ${\approx}10\%$ each in the theory; the larger bicyclic amplitude recovered from the experiment may reflect residual cross-talk between the ring-open and bicyclic patterns. The gradual 2+3 fragment rise is similar in theory and experiment, showing a better temporal match than for isoxazole.

To verify the population dynamics obtained from the low-charge data discussed above, we performed similar fits to the high-charge data (Figure~S7), which show results consistent with the low-charge data and confirm the steady increase in the fragment fractions, plateauing around {3\,ps} for isoxazole and {4\,ps} for oxazole.

The comparison between theoretical predictions and experimental measurements reveals both strengths and limitations of the current UED/computational approach. The qualitative agreement is excellent, with theory correctly predicting the faster dynamics in isoxazole compared to oxazole, the major fragmentation channels, and the overall signatures in the UED patterns. As the structures used for fitting are guided by the theoretical results, it is not surprising that the oxazole shows more complex dynamics than isoxazole at early times with the presence of the bicyclic ring, agreeing with the theory. These structures were not included in the isoxazole fit and hence could not be observed. This is unavoidable as it is impossible for isoxazole to form the bicyclic ring. The similarity in the scattering pattern for the 3-membered ring, ring-open and 2+3 structure along with the large uncertainty in the pattern due to the high internal energy, results in large uncertainty in the weighting of each geometry in the data fits. This uncertainty helps explain some of the discrepancy between the experimental and theoretical fractions.

One key quantitative difference is the time scale. Fragmentation occurs significantly faster in the simulations than in the experiment. This offset suggests physical effects not included in the present semiclassical trajectories. Two are likely relevant: quantum nuclear effects, which would slow fragmentation relative to classical trajectories, and intramolecular vibrational energy redistribution (IVR), which competes with bond breaking by channeling excess energy away from the reaction coordinate. Crucially, this discrepancy is confined to the time scale and does not alter the mechanistic picture.


\section*{Conclusion}
By combining UED experiments with high-level molecular dynamics simulation, we have shown the difference in the photodynamics of the isomers oxazole and isoxazole, revealing the detailed mechanism by which heteroatom positioning determines chemical outcome. The calculations give the most complete picture to date of the transient structures, ring opening, and fragmentation of the two isomers following excitation 
to the $\pi\pi^*$ state at 200 nm. Isoxazole exhibits rapid ring opening along the N-O bond and prompt fragmentation. Oxazole shows more complex dynamics, with the possibility of ring opening at both C-O bonds as well as bicyclic intermediate formation and H atom loss. 
These mechanistic insights have broader implications for understanding photochemistry in related heterocyclic systems. The results suggest that heteroatom positioning can dramatically influence photochemical reactivity through its effects on excited-state potential energy surface topography, conical intersection accessibility, and possible downstream structure formation.

\section{Methods}

\subsection{Ultrafast Electron Diffraction Measurements}
The experiment was performed using the MeV-UED facility at the SLAC National Accelerator Laboratory. The experimental setup has been described in detail elsewhere \cite{Shen2019} so only a brief description will be given here. A simplified diagram of the set-up is shown in Fig.~\ref{fig:scheme}(d). 
The samples of isoxazole and oxazole were purchased from Sigma-Aldrich and used without further purification. The molecules were delivered at room temperature into the high-vacuum interaction region in a continuous-flow gas cell 3 mm in length with 550 μm openings. Both samples were excited at 200 nm with a pulse energy of 10.5~$\mu$J and an approximate pulse duration of 150 fs (FWHM), focused to a spot size of 200 $\mu$m in the flow cell. The sample was probed with a 3.7-MeV electron pulse. Data were acquired using two different electron pulse settings. A `low-charge' data set was recorded with electron bunches with $\approx$1.5$\times$10$^4$ electrons and a fwhm $\approx$150~fs, and a `high-charge' data set was acquired using electron pulses with $\approx$3$\times$10$^4$ electrons and a fwhm $\approx$300~fs. All experimental results shown in the main text are from the `low-charge' data. The `high-charge' data show similar dynamics with lower temporal resolution but better signal-to-noise ratio and are presented in Section~3 of the SI. 

The electron pulse and UV pulse were overlapped at a \ang{1} angle using a mirror with a through hole placed upstream of the flow cell. Both pulses had a repetition rate of 360~Hz. A phosphor screen and a camera recorded the diffracted electrons. The images were collected over several scans with a dwell time of 8~s for isoxazole and 10~s for oxazole at each delay. The order of the delays was randomized for each scan to minimize systematic errors. Time delays between $-$2\,ps and 3\,ps were scanned, with 35 and 38 delay points recorded for isoxazole and oxazole, respectively. Each delay was recorded for a total of 22.5 minutes and 30.7 minutes for isoxazole and oxazole, respectively. The instrument response function was determined to be approximately 300~fs from the fastest changing diffraction feature (see Table S1 in the SI).


\section{Computational Details}

The ground-state geometries of oxazole and isoxazole were optimized at the SA(12)-CASSCF (12,10)/aug-cc-pVDZ level of theory and were used to generate 4000 initial conditions sampled from a Wigner phase-space distribution to incorporate zero-point vibrational motion.\cite{Dunning1989,Kendall1992,Schinke1995,Barbatti2016} Each geometry was excited to adiabatic singlet states within an excitation energy range of 6.6--7.0~eV for isoxazole and 6.4--6.8~eV for oxazole, corresponding to the 200~nm (6.2~eV) excitation pulse used in the experiment. The use of a higher excitation window was based on comparison between the experimental VUV and simulated absorption spectra and is consistent with the fact that CASSCF overestimates excitation energies due to the lack of dynamic correlation.

Nonadiabatic dynamics simulations were performed using the SHARC~3.0 software package at the SA(4)-CASSCF(12,10)/aug-cc-pVDZ level of theory.\cite{Mai2018SHARC,SHARC3.0} Nuclear motion was propagated using Tully's fewest-switches surface-hopping (FSSH) algorithm with velocity--Verlet integration employing a 0.5~fs nuclear timestep, while the electronic equation of motion was integrated using a 0.02~fs timestep for accurate nonadiabatic coupling (NAC) evaluation.\cite{Tully1990} Momentum rescaling upon successful hops was employed to conserve total energy, and frustrated hops were recorded when there was insufficient kinetic energy along the NAC direction. The Granucci--Persico energy-based decoherence correction was employed to reduce overcoherence, and root-tracking ensured consistent state assignment.\cite{Granucci2007}

All excited-state properties including energies, gradients, and nonadiabatic couplings were computed on-the-fly at the SA(4)-CASSCF(12,10)/aug-cc-pVDZ level using OpenMolcas.\cite{OpenMolcas2019} Equal weights were assigned to the lowest four singlet states to ensure stable orbital representation near conical intersections. A total of 106 trajectories for isoxazole and 126 trajectories for oxazole, randomly sampled from the excitation window and initial vibrational phase space, were propagated for 1000~fs. Bond-breaking events were defined when interatomic distances first exceeded their equilibrium value by a factor of 2.5 and remained dissociated for at least 20~fs.

Geometries were classified into structural categories (five-membered ring, bicyclic ring, three-membered azirine ring, ring-opened, and fragment states) using geometrical criteria including ring connectivity analysis and bond distance thresholds, as well as a random forest-based algorithm (see Section~9 of the SI for details). Fragments were grouped by heavy-atom partitioning (2+3 for two-body or 2+2+1 for three-body) due to UED's limited sensitivity to hydrogen atoms. Representative conical intersections were optimized using the penalty-function minimization method.\cite{Levine2008} Attachment--detachment densities were computed to characterize the electronic nature of excited states.\cite{HeadGordon1995}

Elastic scattering patterns were simulated via the independent-atom model (IAM) from the trajectory geometries, ensemble-averaged to yield $\Delta I/I$ profiles, and temporally convolved with a 300~fs Gaussian instrument response function to match the experimental conditions. 
Theoretical basis patterns were averaged for each structural class and used to fit experimental diffraction data through least-squares minimization. Further details can be found in the Supporting Information.

\section*{Data Availability Statement}

The experimental UED data and trajectory geometries that support the findings of this study are available from the corresponding authors upon request. The SHARC and OpenMolcas software packages used for nonadiabatic dynamics simulations are publicly available.

\section*{Supporting Information}

The Supporting Information is available free of charge. It contains additional details on the experiment; high-charge experimental data; instrument response function determination; additional theory results and trajectory analysis; detailed fragment pathways; and details on the machine learning classification.

\section*{Acknowledgements}
The experiment was carried out at the SLAC MeV-UED user facility, operated as part of the Linac Coherent Light Source at the SLAC National Accelerator Laboratory and supported by the U.S.~Department of Energy, Office of Science, Office of Basic Energy Sciences under Contract No.~DE-AC02-76SF00515, during beamtime U110. We thank the technical and scientific staff at SLAC for their hospitality and support before and during the beamtime. Other authors are funded through the Chemical Sciences, Geosciences, and Biosciences Division, Office of Basic Energy Sciences, Office of Science, U.S.~Department of Energy under grant nos.~DE-SC-17300 (A.G.S.), DE-FG02-86ER13491 (S.K.S., S.G., V.K., H.V.S.L., T.T.N., A.R., J.S., D.R.); and DE-SC0020276 (M.C., J.P.F.N.); National Science Foundation grant no.~PHYS-2409365 (A.S.V.); Strategic Priority Research Program of the Chinese Academy of Sciences, Grant No.~XDB34020000, 12450404, the CAS Pioneer Hundred Talents Program, and the USTC Research Funds of the Double First-Class Initiative, YD2030002020 (E.W.). The machine learning aspect of this work was supported by a GRIPex award from Kansas State University. Computational resources were provided by the Beocat Research Cluster at Kansas State University

\bibliography{UED_corrected} 

\end{document}